**Title**

**Physical Vapor Deposition of High Mobility P-type Tellurium and its Applications for Gate-tunable van der Waals PN Photodiodes**

*Tianyi Huang[1], Sen Lin[2], Jingyi Zou[2], Zexiao Wang[1], Yibai Zhong[1], Jingwei Li[2], Ruixuan Wang[2], Han Wang[3], Qing Li[2], Min Xu[2, 4], Sheng Shen[1,2,5]\*, Xu Zhang[1,2]\**

[1] Department of Mechanical Engineering, Carnegie Mellon University, Pittsburgh, PA, 15213, USA

[2] Department of Electrical and Computer Engineering, Carnegie Mellon University, Pittsburgh, PA, 15213, USA

[3] Department of Electrical and Electronic Engineering, The University of Hong Kong, Hong Kong

[4] Ray and Stephanie Lane Computational Biology Department, School of Computer Science, Carnegie Mellon University, Pittsburgh, PA, 15213, USA

[5] Department of Materials Science and Engineering, Carnegie Mellon University, Pittsburgh, PA, 15213, USA

Abstract

Recently tellurium (Te) has attracted resurgent interests due to its p-type characteristics and outstanding ambient environmental stability. Here we present a substrate engineering based physical vapor deposition method to synthesize high quality Te nanoflakes and achieved a field-effect hole mobility of 1500 cm$^2$/Vs, which is, to the best of our knowledge, the highest among the existing synthesized van der Waals p-type semiconductors. The high mobility Te enables the fabrication of Te/MoS$_2$ pn diodes with highly gate-tunable electronic and optoelectronic characteristics. The Te/MoS$_2$ heterostructure can be used as a visible range photodetector with a current responsivity up to 630 A/W, which is about one order of magnitude higher than the one achieved using p-type Si-MoS$_2$ PN photodiodes. The photo response of the Te/MoS$_2$ heterojunction also exhibits strong gate tunability due to their ultrathin thickness and unique band structures. The successful synthesis of high mobility Te and the enabled Te/MoS$_2$ photodiodes show promise for the development of highly tunable and ultrathin photodetectors.

## 1. Introduction

The PN diode, typically made of a junction between p-type and n-type semiconductors, is one of the most fundamental building blocks of electronics[1], with a wealth of applications in photovoltaics[2], integrated circuits[3,4] and photodetectors[5]. Since the first successful isolation of graphene[6], two-dimensional (2D) van der Waals semiconductors[7] have emerged as promising candidates that could potentially transform traditional silicon-based electronics with new form factors, such as flexible, ultralight and/or transparent circuits in the post-Moore era[8,9]. Despite substantial advances in the synthesis and engineering of n-type 2D semiconductors, superior and stable p-type 2D semiconductors are much less explored[10]. It hindered the development of a diverse portfolio of 2D PN diodes, in which p-type semiconductors are essential components. Tellurium (Te) is an elemental p-type van der Waals semiconductor, possessing 1D helical chains of covalent bonds and the adjacent chains packed via van der Waals interactions in a hexagonal lattice[11-15]. Studies on Te attracts considerable resurgent interest especially due to its excellent environmental stability[11] which is a clear advantage compared with black phosphorous (BP) which is one of the most extensively studied high mobility p-type semiconductors[16,17].

Due to its unique chiral-chain structure and relatively strong bonding, it is extremely difficult, if not impossible, to mechanically exfoliate Te flakes from its bulk form like other van der Waals materials[18]. Wang et al. used a hydrothermal method to synthesize Te nanoflakes through the reduction of sodium tellurite (Na$_2$TeO$_3$) and achieved hole mobility of 700 cm$^2$V$^{-1}$s$^{-1}$ [11]. However, the involvement of organic solvents often introduces impurities and surface trap states during the synthesis and suffers from limited scalability. As an alternative technique, hydrogen-assisted chemical vapor deposition (CVD) has also been demonstrated to synthesize 2D Te flakes on mica substrates[19]. Zhao et al. also showed a chemical vapor transport method to synthesize 2D Te nanoflakes with thickness down to 70 nm and with mobility of 379 cm$^2$/Vs [20]. Yang et al. significantly improved the hole mobility of synthesized Te nanobelts up to 1370 cm$^2$/Vs using a CVD method through high temperature reduction of TiO$_2$[21]. Meanwhile, Zhou et al. showed Te nanostructures can be obtained by molecular beam epitaxy (MBE) at low temperature (≤120 °C) and exhibited field-effect mobilities up to 707 $cm^2V^{-1}s^{-1}$[22]. Recently, physical vapor deposition (PVD) has emerged as another promising strategy to synthesize Te in a scalable manner[23–27]. Compared with CVD approaches, PVD method typically allows relatively low synthesis temperatures, and therefore a wider range of allowed substrates, as well as produce minimal hazardous byproducts. For example, Zhao et al. demonstrated that thermal evaporation at cryogenic temperatures can produce wafer-scale Te thin films with an effective hole mobility of ~35 $cm^2V^{-1}s^{-1}$[27]. Controllable unbalanced magnetron sputtering can also be used to obtain ultrathin Te films with carrier mobility up to 19 $cm^2V^{-1}s^{-1}$[23]. Despite these recent progresses, the mobility of as-synthesized Te still has plenty of room to improve.

In this work, we developed a physical vapor deposition method with substrate engineering to synthesize high quality Te nanoflakes and achieved a field-effect hole mobility of 1500 $cm^2/Vs$, which is, to the best of our knowledge, the record high performance among existing synthesized Te and other alternative 2D p-type semiconductors. Substrate engineering is a widely used strategy in CVD approaches to facilitate oriented growth of 2D materials. Here, by introducing atomically flat hBN as growth substrates in PVD systems, we found that p-type Te with ultrahigh hole mobility can be synthesized. To demonstrate the high quality of as-synthesized Te, we built a Te-$MoS_2$ 2D heterojunction as a gate-tunable PN diode. The rectification ratio of the Te-$MoS_2$ diode can reach 4000 and can be electrostatically tuned through gating. Under white light illumination, the Te-$MoS_2$ as be used as a photodetector with current responsivity up to 630 A/W, which is about one order of magnitude higher than the one achieved using p-type Si-$MoS_2$ PN photodiodes. Importantly, under different gate bias, the responsivity of the Te-$MoS_2$ diode exhibits different dependance on light power density. The tunable photoresponse of the Te-$MoS_2$ diode can potentially open up new opportunities in building ultralight and multifunctional optoelectronic devices.

## 2. Results and Discussion

Here, ultrathin Te was synthesized by the PVD approach as illustrated in **Figure 1**a. hBN flakes were mechanically exfoliated onto a piece of silicon wafer capped with 285 nm $SiO_2$, serving as the atomically smooth growth substrate. Bulk Te (from Sigma Aldrich) was used as the precursor placed in a quartz boat and the growth substrate was loaded in the downstream region about 9 inches away from the Te precursor (Fig. 1a). A mixture of argon and hydrogen was used as the carrier gas to deliver precursor vapor for synthesis. Detailed experimental parameters can be found in the Experimental Section. Figure 1b shows a typical optical image of as-synthesized Te flakes on hBN substrates. Raman characterization shows two peaks at 120 cm$^{-1}$ of A$_1$ mode and 140 cm$^{-1}$ of E$_2$ mode (Fig. 1c), which are consistent with the literature[21,28].

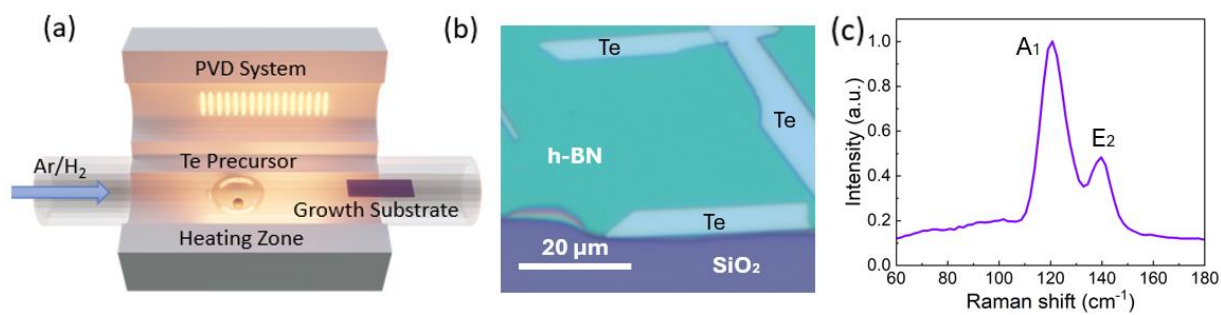

**Figure 1.** (a) Schematic of PVD process for 2D Te thin flakes growth. (b) Optical images of the as-synthesized Te flakes. (c) Raman spectrum of the PVD-grown Te flakes.

Field effect transistors (FETs) with four-point contacts were then fabricated based on the PVD-synthesized Te to characterize its transport properties (Figs. 2a and 2b). Gold (Au) was used as the source/drain electrodes and heavily doped silicon substrate was used as the backgate terminal. The thickness of the Te channel was determined by atomic force microscopy (AFM, Fig. 2c). The Te channel has smooth surface and uniform thickness of about 30 nm, as shown in Figure 2d. Quasi-static IV characterization of the Te FET was carried out by using semiconductor parameter analyzer (B1500). IV transfer curves in both linear and logarithm scale were shown in Figure 2e, under drain bias V$_{ds}$ of 0.5 V and a backgate bias swept from -100 V to 100 V. The IV measurement confirms the p-type behavior of Te for V$_g$ < 30V. The transfer curve of the Te FET exhibits ambipolar characteristics to certain degree, indicating relatively small bandgap of Te, which is also consistent with literature[11,21,29]. The on/off ratio of this FET is around 44. The field-effect hole mobility can be extracted by the following equation.

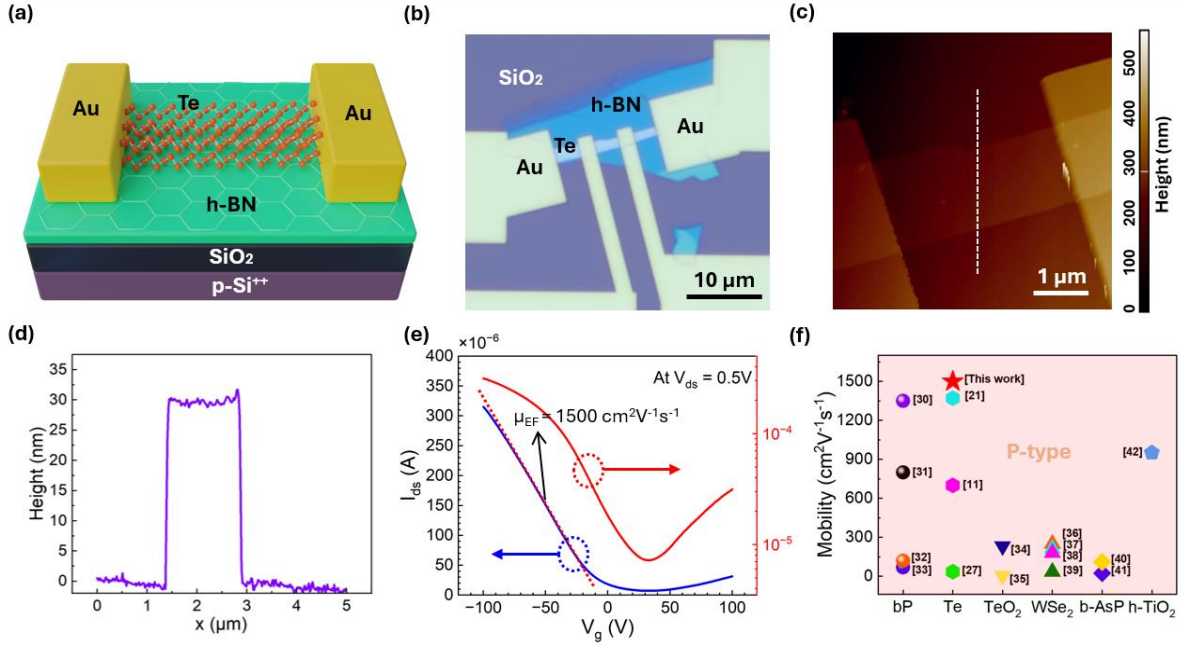

**Figure 2.** (a) Schematic of Te FETs. (b) An optical image of the as-fabricated four-point Te FET. (c) An AFM scanning image of the Te FET. (d) AFM height profile of the Te FET channel (along white dash line). (e) IV transfer curve of the Te FET ($V_{ds}$ = 0.5 V) in both linear (blue line) and logarithmic (red line) scales. (f) Measured field-effect mobility of the Te FET, benchmarked with reported traditional p-type 2D semiconductors.

$$\mu_{EF} = \frac{\Delta I_{ds}}{\Delta V_g} \times \frac{L^2}{W \times C_{gate} \times V_{ds}} \qquad (1)$$

, where $I_{ds}$ is the channel current, $V_g$ is the backgate bias, L is the channel length, W is the channel width, $C_{gate}$ is the capacitance of the $SiO_2$ dielectric layer, and $V_{ds}$ is the drain bias. The hole mobility of the Te FET was estimated to be 1500 $cm^2V^{-1}s^{-1}$, which is, to the best of our knowledge, the highest field-effect mobility in p-type 2D semiconductors. We benchmarked our result with state-of-the-art p-type semiconductors in Fig. 2f.

The synthesis of high-mobility p-type Te flakes also makes it possible to build van der Waals PN heterojunction as photodiodes. The high mobility of Te is particularly beneficial in improving the efficiency of Te-based PN photodiodes. We chose $MoS_2$ as the N-type semiconductor to form the cathode terminal. The Te-$MoS_2$ van der Waals heterojunction devices are fabricated by transferring mechanically exfoliated $MoS_2$ few-layer flakes on top of the as-synthesized Te flake. Figure 3a shows the schematic illustration of the Te-$MoS_2$ van der Waals heterojunction device. Gold was used as the contacts to both terminals of the Te-$MoS_2$ diode and the heavily doped silicon substrate was used as the back gate terminal. Figure 3b is one representative SEM image of the heterojunction device after fabrication. The IV characteristics of the Te-MoS$_2$ diode in the dark were first measured using a semiconductor analyzer (B1500A). By sweeping the bias applied between the anode (Te) and cathode (MoS$_2$) from -2V to +2V, the channel current I$_D$ exhibits strong rectification behaviors. The rectification ratio is highly tunable by changing the gate bias. Figure 3d shows the calculated rectification ratios obtained from the *I-V* curves at $V_D$ = 2V at different gate biases. With the gate voltages varying from -30V to +20V, the rectification ratio increases from 128 to 3644. The current rectification can be explained by the energy band diagram in Figs. 4a-4b. Electrons in the n-type MoS$_2$ see a higher energy barrier at the Te/MoS$_2$ interface under reverse bias. Under forward bias, this energy barrier is reduced, and the current is mainly from the thermionic emission of electrons from MoS$_2$ to Te. The diffusion of these electrons inside the Te gives rise to the forward current. The rectification ratio exhibits a strong gate bias dependence. The forward current is more sensitively dependent on the gate bias. This is because, under forward bias, the thermionic emission of electrons depends on the electron density near the interface (Figs. 4c-4d). By increasing the gate bias, doping level of MoS$_2$ and thus the electron density can be enhanced.

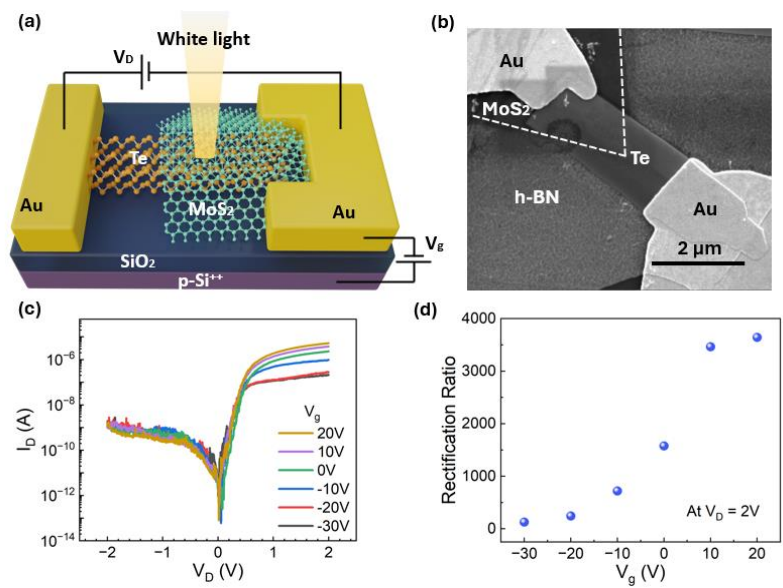

**Figure 3**. (a) Schematic of the Te-MoS$_2$ van der Waals PN heterojunction based photodiode. (b) SEM image of one representative Te-MoS$_2$ heterojunction diode. The shape of the MoS$_2$ flake is indicated by the white dash line. (c) Measured channel current I$_D$ (on a logarithmic scale) as a function of channel bias V$_D$ at various gate voltages. (d) Calculated rectification ratio as a function of gate voltage V$_g$ at V$_D$ = 2 V.

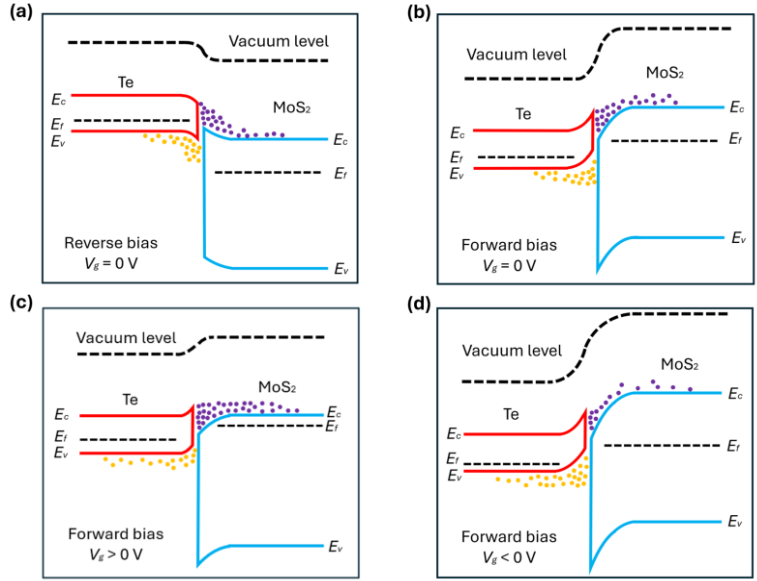

**Figure 4.** Band structure of the Te-MoS$_2$ van der Waals PN heterojunction under a reversed (a) and forward (b) voltage bias without applying gate voltage. The majority carriers in MoS$_2$, i.e. electrons, are indicated as purple color dots. The majority carriers in Te, i.e. holes, are indicated as yellow color dots. Electrons in MoS$_2$ see different energy barriers at the MoS$_2$/Te interface under forward and reverse biases, resulting a current rectification phenomenon. (c) Band structure of the heterojunction under a forward voltage bias with a positive gate voltage applied. (d) Band structure of the heterojunction under a forward voltage bias with a negative gate voltage applied.

The ultrathin nature of the Te-MoS$_2$ heterostructure allows gate tunability of its diode IV characteristics. It is a unique feature compared to conventional silicon based PN diodes, for which the electrical properties are essentially fixed and determined by their pre-defined doping profiles. In order to exploit the gate-tunability in photodetectors, we measured the photoresponses of the Te-MoS$_2$ diode under different gating biases. A white light optical beam illuminates the Te-MoS$_2$ device in normal incidence. The detailed spectrum of the incident light was measured and shown in the supplementary information. We first measured the channel current I$_D$ at zero gate bias by sweeping the channel bias V$_D$ from -0.4 V to 0.4 V and keeping the V$_g$ to be zero, under incident light intensity ranging from 13.6 mW/cm$^2$ to 83.2 mW/cm$^2$ (measured by PM100D from Thorlabs). Then, the photon current $I_{ph}$ can be obtained by extracting the dark current. The current responsivity can be calculated as follows.

$$R = \frac{I_{ph}}{|P \times A|} \qquad (3)$$

where $I_{ph}$ is the photocurrent, $P$ is the incident light intensity and $A$ is the overlapping area of the Te/MoS$_2$ heterojunction. As shown in Figure 5b, the photocurrent linearly increases as the incident light power increases. We note that the forward biasing regime gives a higher responsivity than the reverse regime. Therefore, the current responsivity in Fig. 5 were measured under forward bias. The current responsivity under zero gate biasing is essentially independent of the incident light power in the range from 13.6 mW/cm$^2$ to 83.2 mW/cm$^2$ (Fig. 5c). The average responsivity is about 593 A/W in this range of light power. This can be explained by the energy band diagram showin Fig. 5d. MoS$_2$ is the photosenistive layer due to its bandgap matching with the incident light frequency. The incoming photons generates electron hole pairs in the MoS$_2$ and the electric field inside the depletion region is in favor of hole drift from the inteface to the MoS$_2$ side. It contributes to a photocurrent that is in the same direction as the dark current of Te/MoS$_2$ under forward bias. The unique Te/MoS$_2$ band structure makes it different from the conventional silicon pn junction based photodetectors, in which the photocurrent is in the opposite direction of dark current and its maximum responsivity occures under reverse bias. The atomic thickness of MoS$_2$ also minimizes the photocarrier recombination rate, which benefits the enhancement of the photocurrent and responsivity.

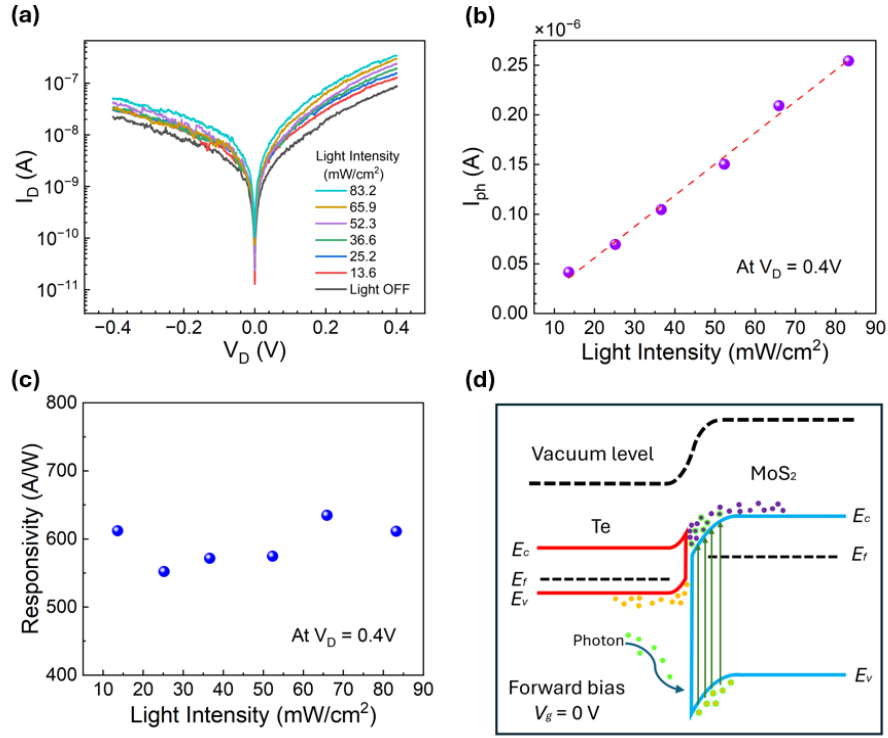

**Figure 5.** (a) Measured channel current $I_D$ (on a logarithmic scale) as a function of channel bias of the Te-MoS$_2$ heterojunction device under different light intensity. The gate bias is zero for all curves. (b) Photocurrent as a function of light intensity at channel bias $V_D = 0.4V$. (c)

Calculated current responsivity as a function of light intensity at channel bias $V_D = 0.4V$. (d) Band structure of the heterojunction photodiode.

Next, the photocurrent of the Te-MoS$_2$ diode was measured at different gate biases in order to investigate its gate tunability. Figure 6a shows the *I-V* curves of the Te-MoS$_2$ photodiode under 0.4 V constant forward bias by sweeping the gate voltage $V_g$ from -20 V to +20 V. From the photocurrent measurement, we found that the current responsivity R of Te-MoS$_2$ photodiode increases as the gate bias increases for all tested light intensities (Fig. 6b). It worths noting that the dependence of responsivity R on incident light intensity is different under different gate biases. At $V_g$ = -15V, the responsivity R increases linearly as the light intensity increases (Fig. 6c). However, at $V_g$ = -10V and above, the responsivity exhibits minimum dependence on the ligth intensity (Fig. 6d). It indicates the existence of traps states in the Te/MoS$_2$ heterostructure. When Vg = -15V, most of the these trap states are empty and a significant fraction of photo-generated carriers under low illumination is trapped and cannot contribute to the photocurrent. Only under high incident light intensity, more of the trapped states are occupied and a higher photocurrent can be observed. At Vg = -10V, the electron density in MoS$_2$ increases and majorty of the trap states are occupied. The impact of trap states on the responsivity becomes much less significant. This is also consistent with the measurement result of Fig. 5c.

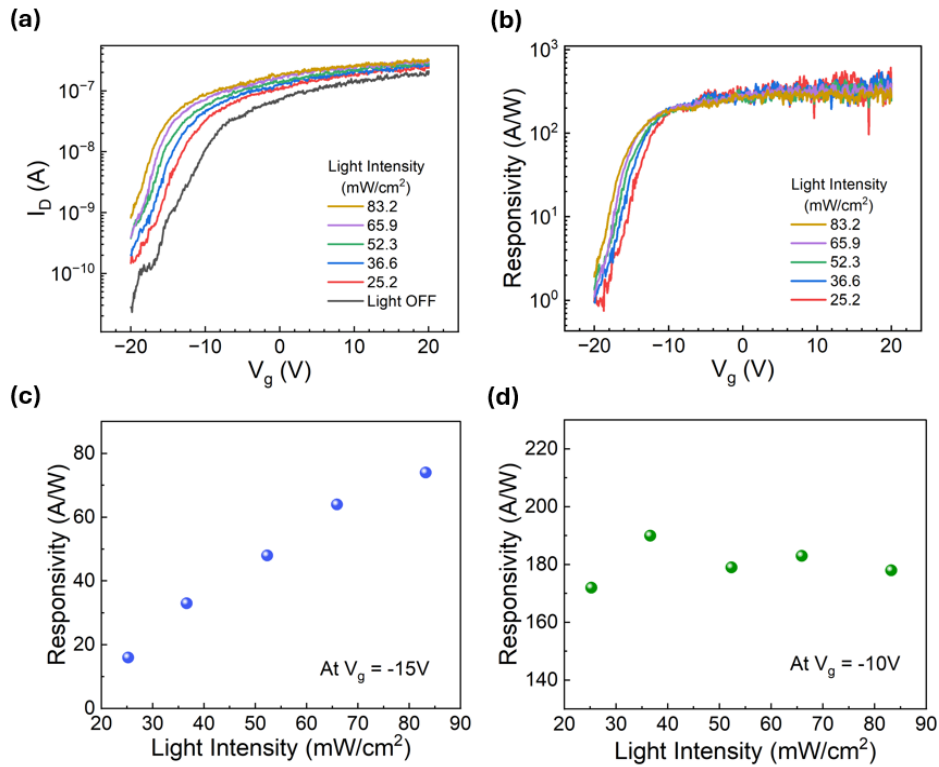

**Figure 6.** (a) Measured channel current $I_D$ as a function of gate voltage $V_g$ under various light intensity. The Te-MoS$_2$ diode was biased under with a forward bias $V_D$ of 0.4V. (b) Calculated responsivity as a function of gate voltage $V_g$ obtained from the IV curves in (a). Responsivity as a function of light intensity under gate voltage of -15 V (c) and -10 V (d).

## 3. Conclusion

In conclusion, we presented a substrated engineering based PVD synthesis approach which produces Te with a record high field-effect hole mobility of 1500 cm$^2$/Vs among state of the art synthesized 2D p-type semiconductors. We demontrated Te/MoS$_2$ pn heterojunction based on the PVD-synthesized Te and it exhibits strong gate tunablity. In the visible range, the Te/MoS$_2$ exhibits photodetection capablity with a current responsivity up to 630 A/W, which is about one order of magnitude higher than the one achieved using p-type Si-MoS$_2$ PN photodiodes. The photo response of the Te/MoS$_2$ photodiode also shows strong gate tunability and distinct optical characteristics that are different from conventional silicon-based photodetectors. The demonstrated Te/MoS$_2$ photodetectors create new possibilities in the development of ultralight and highly tunable optoelectronic devices.

## 4. Experimental Methods

*PVD synthesis of 2D tellurium thin flake:*

Large h-BN flakes (purchased from 2D Semiconductor) were mechanically exfoliated on a 2 × 0.5 inch silicon substrate initially. During the PVD process, a piece of bulk Te precursor (from Sigma Aldrich) was placed in a quartz boat located at the center of the PVD quartz tube as the evaporation source. The silicon substrates deposited with h-BN flakes were then placed around 9 inches from the Te source downstream (Fig .1a). After the loading process, 900 sccm Ar and 100 sccm H$_2$ were first flowed for about 4 minutes to purge the PVD tube. Then, a Linderberg Blue M furnace (from Thermo Scientific) was used to heat up the tube to 660 ℃ from room temperature (22℃) with 40℃/min ramp rate. A flow of 35 sccm Ar is introduced from one end of the tube during this heating process and the Ar gas carried the vaporized Te onto the substrate for synthesis. The Te vapor condensed on the surface of the substrate with low temperature and formed nanoflakes. After reaching 660 ℃, the flow of Ar gas was stopped and the PVD tube is cooled immediately. The thickness of most Te flakes grown by this PVD process ranging from 10 nm to 100s nm.

*Device Fabrication:*

For the fabricaiton of Te transistors, the source and drain electrodes were defined by electron beam lithography system (Elionix ELS-G100). Then, a layer of 50 nm gold (Au) was deposited by Lesker PVD-75 electron beam evaporation (EBE) at 1 Å/s, followed by a liftoff process. For the Te-MoS$_2$ heterojunction devices, the MoS$_2$ (from 2D Semiconductors) flakes were mechanically exfoliated onto poly(methyl methacrylate)/polyvinylpyrrolidone (PMMA/PVP), which were spin-coated on a silicon chip in advance. Next, the water soluble PVP layer was dissolved by DI water rendering the PMMA layer with MoS$_2$ flakes on top separated from the silicon chip and floating on the DI water. This PMMA thin film was then transferred onto the top of Te flakes under microscope in an aligned way. The PMMA film was subsequently dissolved by acetone. Contact electrodes were then made using the same process as used for the fabrication of Te transistors.

**Supporting Information**

Supporting Information is available from the Wiley Online Library or from the author.

**Acknowledgements**

This work was financed in part by National Science Foundation (Grant number: NSF ECCS-2239822), a Grant (contract C000080565) from the Commonwealth of Pennsylvania, Department of Community and Economic Development, and support from Carnegie Mellon University. TH, SL, JZ, YZ, SS and XZ also acknowledge the support from the CMU Claire & John Bertucci Nanotechnology Laboratory.

Received: ((will be filled in by the editorial staff))
Revised: ((will be filled in by the editorial staff))
Published online: ((will be filled in by the editorial staff))

**Title**

Physical Vapor Deposition of High Mobility P-type Tellurium and its Applications for Gate-tunable van der Waals PN Photodiodes

ToC figure ((Please choose one size: 55 mm broad × 50 mm high **or** 110 mm broad × 20 mm high. Please do not use any other dimensions))

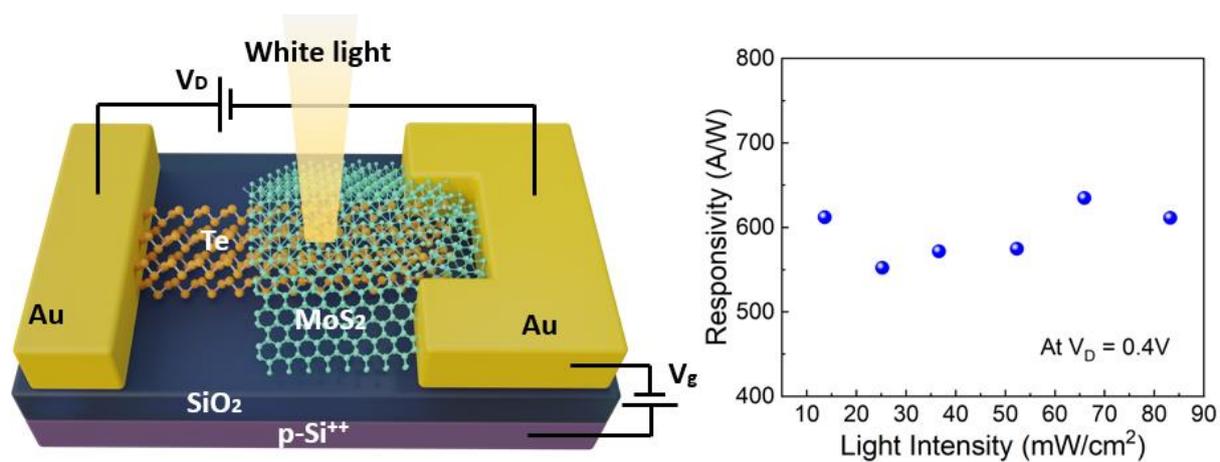

# WILEY-VCH

## Supporting Information

Title: Physical Vapor Deposition of High Mobility P-type Tellurium and its Applications for Gate-tunable van der Waals PN Photodiodes

*Author(s), and Corresponding Author(s)** ((write out full first and last names))

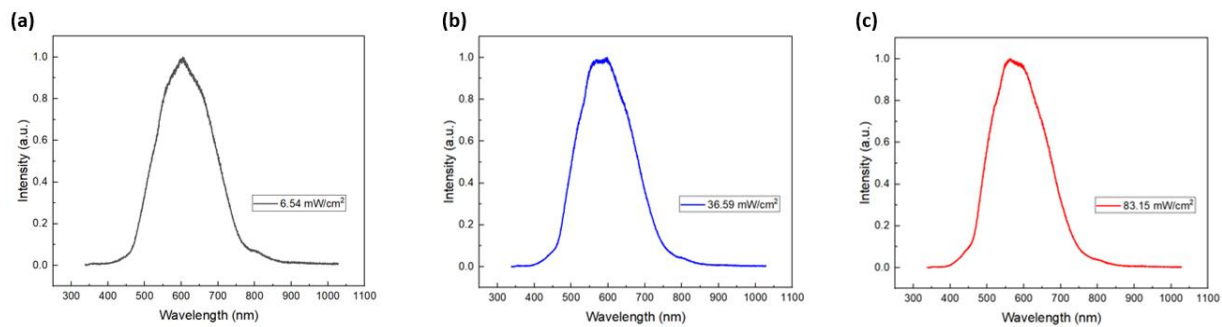

**Figure S1.** Spectrum of the light source under (a) 6.45 mW/cm$^2$, (b) 36.59 mW/cm$^2$ and (c) 83.15 mW/cm$^2$